\documentstyle[epsf,aps,prl]{revtex}

\begin{document}
\draft
\title{Thermodynamic functions within the van Hove BCS model: Symmetry mixing
effects.}
\author{D. Quesada, R. Pe\~{n}a, andC. Trallero-Giner}
\address{Departamento de F\'{\i }sica Te\'{o}rica, Universidad de La Habana, \\
La Habana 10 400, Cuba}
\maketitle

\begin{abstract}
All the new layer perovskite superconductors seem to show a phenomenon of
symmetry mixing with repect to the order parameter. An analysis of the
different alternative of mixing and how far could them be presented is
carried out. For the particular case of s+id symmetry of the gap, the
temperature dependence of the specific heat ($C_{es}$) and the thermodynamic
critical magnetic field ($H_{c}$) are calculated. A double peak transition
is observed on $C_{es}(T)$ in the mixed regime while the single peak
behavior is recovered for a purely symmetric state ( s or d). $C_{es}$
presents a quadratic law at low tempeartures for a d-wave gap and for the
s-wave one the typical exponential attenuation. The temperature dependence
of $H_{c}$ shows a clear phase transition of second order at temperatures
where the d-wave component becomes negligible. A comparison with other
results and experiments is done.
\end{abstract}



\section{Introduction.}

The discovery of the high temperature superconductivity (HTSC) in layered
perovskites has stimulated many theoretical studies that range from
phenomenological approaches to microscopic models. Some of them follow the
traditional Fermi liquid picture with a BCS \cite{schri1} or Eliashberg \cite
{eliash} formulations, while others, often focusing on the new electronic
fluids, use to follow the so called marginal \cite{varma} and nearly
antiferromagnetic Fermi liquids \cite{monien}. However, up to date a fully
consistent interpretation of the large amount of existing experimental data
is not found in the literature and further efforts are needed for the
understanding of the pairing mechanism in cuprate perovskites. In this
context, it is important to compare the thermodynamic functions of these new
compounds and the old layered superconductors with low critical
temperatures, looking for similarities and differences between them.

The cuprate perovskites are layered materials with highly anisotropic
properties which show an almost dispersionless electronic bands with flat
sections in the vicinity of the Fermi level ($\varepsilon _{F}$). This leads
to the appearance of the so called van Hove singularities (vHs) in the
electronic density of states (EDOS) \cite{shen}. On the other hand, in these
compounds the possibility of gap symmetries different from the s-wave one is
a matter of debate up to now \cite{klemm,wenger}. One of the first attempts
to explain the great variety of behaviors experimentally observed was the
BCS model within the vHs scenario using the s-wave symmetry for the order
parameter \cite{bok,tsuei2,get,davidq}. Nevertheless, several experiments,
such as tricrystal corner junction tunnelling in $YBa_{2}Cu_{3}O_{7-\delta }$
\cite{tsuei1} show strong evidence that the gap symmetry was dominated by a $%
d_{x^{2}-y^{2}}$ component. Following these ideas and in the framework of
the BCS model, in Refs.\cite{tsuei3,maki1}, the temperature dependence of
the gap, the ratio $2\Delta (0)/k_{B}T_{c},$ and the specific heat $C_{es}$
as a function of temperature were calculated. While this approach seems to
yield results more close to the observed experimental behavior of those
magnitudes, several reports ( see \cite{klemm} and references therein) are
very hard to explain in terms of purely symmetry states. In this sense,
recent experiments on c axis tunneling in the $Pb-YBCO$ system \cite{ghosh}
(and references therein) reveals that the superconducting order parameter
should have a small s wave contribution in addition to the d wave one. With
that aim, in Refs. \cite{ghosh,carbote,betouras1,liu} the phase diagram for
the $s+d$ and $s+id$ gaps as a function of the filling factor and the
driving parameter $V_{d}/V_{s}$ were studied in the framework of the BCS
model. The purpose of this paper is to study the effects of the symmetry
mixing of the gap on the thermodynamic functions as the electronic specific
heat $C_{es}$ and the critical magnetic field $H_{c}$ within the weak
coupling van Hove BCS (v-BCS) model.

The paper is organized as follows. In Sec. 2 the theoretical background and
gap symmetries are introduced within the van Hove singularity scenario. In
Sec. 3 the thermodynamic functions of the superconducting material as a
function of temperature are obtained. Finally, in Sec. 4 the main
conclusions of the paper will be given.

\section{Toy model for a mixture of gap symmetries.}

In agreement with the layered structure of the new superconducting materials
we start from a tight binding formulation of the band structure with the
following dispersion law: $\varepsilon _{\vec{k}}=-2t(cosk_{x}a+cosk_{y}a)$,
where $t$ is the coupling parameter between the Cu and O atoms, $k_{x}$ and $%
k_{y}$ are the components of the wavevector, and $a$ is the lattice
constant. This band model leads to a density of states similar to that
obtained in Ref.\cite{davidq}. Furthermore, we have considered the weak
coupling limit of the BCS theory including gap symmetries different from the
often used s- or d- waves as well as its possible mixing. We are discussing
mainly the mixed $s+d$ and $s+id$ cases which seem to be closest to
experimental results. The tetragonal lattice symmetry assumed in the band
structure will impose restrictions on the possible forms of $\Delta _{\vec{k}%
}$ and for the gap the following expression are assumed \cite{klemm,wenger}: 
\begin{eqnarray}
\Delta _{\vec{k}}(T)=\Delta _{\vec{k}}^{s}(T)+\exp (i\theta )\Delta _{\vec{k}%
}^{d}(T),  \label{gap}
\end{eqnarray}
with 
\begin{eqnarray}
\Delta _{\vec{k}}^{s}(T) &=&\Delta _{s}(0)g(T/T_{c}^{s}),  \label{gap-s} \\
\Delta _{\vec{k}}^{d}(T)&=&%
\Delta_{d}(0)(cos(k_{x}a)-cos(k_{y}a))g(T/T_{c}^{d}).  \label{gap-d}
\end{eqnarray}
The term $\exp (i\theta )$ in Eq.(\ref{gap}) enables one to fix the relative
phase between the s and d symmetries. For $\theta =0$ we are dealing with
the $s+d$ mixing for which the order parameter is a real quantity, while for 
$\theta =\pi /2$ the $s+id$ mixing a complex order parameter is presented.
On the other hand, two different critical temperatures, $T_{c}^{s}$ and $%
T_{c}^{d}$ should appear in agreement with the respective symmetries of the
gap, so that the critical temperature of the whole superconductor will be
the maximum among these two values $(T_{c}={\bf max}(T_{c}^{s},T_{c}^{d}))$.
Which of the two symmetries is responsible for the critical temperature
depends on many factors \cite{klemm,carbote,betouras1} as for example, at
higher filling values the $d_{x^{2}-y^{2}}$ gap is favored, while for lower
filling factor we have a pure s-gap \cite{carbote}. Irrespective of the
conditions favoring one or another gap symmetry mixing we are interested on
the influence of this on the thermodynamic functions. With this purpose and
to simulate how the material evolves from s to d-wave, the following
equations for $T_{c}$ are used: 
\begin{eqnarray}
\frac{k_{B}T_{c}^{s}(c)}{2t} &=&\frac{1}{\exp [(c-0.3)/0.04]+1}, \\
\frac{k_{B}T_{c}^{d}(c)}{2t} &=&1-\frac{1}{\exp [(c-0.5)/0.06]+1},
\end{eqnarray}
where $c$ is a parameter that drives the amount of d-wave component and a
suitable estimation for this parameter could be the ratio $V_{d}/V_{s}$. The
values $c=0$ and $c=1$ correspond to the nearly s- and d-wave gaps,
respectively. This model is consistent with results obtained in Refs.\cite
{carbote,liu} where three regions are well defined. Two of them have a
dominant gap symmetry, s or d and the third one a mixture of these two
symmetries. On the other hand, and irrespective of the gap symmetry assumed,
we denote by $g(T/T_{c})$ the temperature dependence of the gap which with a
good accuracy can be represented by the following relation \cite{dorbolo}: 
\begin{eqnarray}
g(\frac{T}{T_{c}})=\tanh \left[ 2\ \sqrt{\frac{T_{c}-T}{T}}\right] .
\label{gaptemp}
\end{eqnarray}
It is worth noticing that the function $g(T/T_{c})$ differs slightly from
the correct BCS dependence and the slope at $T_{c}$ is about 1.6 times the
true one.

Since the thermodynamics of any superconductor is governed by the concrete
topology of the quasiparticle excitation energy, in Fig.1 the contour lines
generated by the surface $E_{\vec{k}}=\sqrt{\varepsilon _{\vec{k}%
}^{2}+|\Delta _{\vec{k}}|^{2}}$ are plotted for the s-wave (upper left) and
d-wave (upper right) pure symmetry cases as well as the $s+d$-wave (bottom
left) and $s+id$-wave (bottom right) mixing at zero temperature. The contour
lines of the symmetry mixing cases are plotted considering the d-wave
component as the dominant one and assuming $T_{c}^{s}=0.7T_{c}^{d}$. In all
cases, $E_{\vec{k}}$ is expressed in units of $2t$ as a function of the
dimensionless wave vectors components $X=k_{x}a$ and $Y=k_{y}a$. As can be
seen from these pictures, the function $E_{\vec{k}}$ does not have zeros in
the pure s-wave case while for the d-symmetry isolated zeros at ($\pm \pi
/2,\pm \pi /2$) are present. These zeros lead to qualitatively different
results in the behavior of thermodynamic quantities at low temperatures.
While for a s-wave gap an exponential attenuation in $C_{es}$ is expected
in agreement with the activation character of the quasiparticle spectrum,
for the d-wave case a power law follows, due to the absence of any
threshold. Concerning the mixing of symmetries, it is clear that the $s+d$
and $s+id$ gaps should lead to a different character form of the
thermodynamic functions. Since the symmetry of the contours associated with
the $s+d$ mixing is lower than the tetragonal one this kind of mixture is
more suitable for lattice with a small ortorhombic distortion, as for
example in the $YBa_{2}Cu_{3}O_{7-y}$ system.

\section{Thermodynamic Functions.}

From experimental point of view a double peak transition in the specific
heat ($C_{es}$)has been observed for YBCO \cite{choy,butera,junod}. In Ref.%
\cite{butera} a double step at 88 K and 92 K were reported by adiabatic
continuum method. Moreover, in Refs.\cite{choy,junod} a review of the
experimental evidences about the
double transition in $C_{es}$ for $ReBa_{2}Cu_{3}O_{7}$ ($Re=Y,Gd$) system 
has been presented. These experimental facts
are in correspondence with the mixed gap scenario. Starting from the above
motivation in this section we are dealing with the specific heat in the
superconducting state and the thermodynamic critical magnetic field taking
into account the gap symmetry mixing.

The electronic specific heat in the superconducting state can be written as: 
\begin{equation}
C_{es}(T)=\frac{2}{k_{B}T^{2}}\sum_{\vec{k}}\ f(E_{\vec{k}})[1-f(E_{\vec{k}%
})][E_{\vec{k}}^{2}-\frac{1}{2}T\frac{d|\Delta _{\vec{k}}(T)|^{2}}{dT}].
\end{equation}
$f(E_{\vec{k}})$ being the Fermi-Dirac distribution function. The derivative
of the square of the gap in the case of a mixture of symmetries can be cast
into: 
\begin{eqnarray}
\frac{d|\Delta _{\vec{k}}(T)|^{2}}{dT}=\frac{d\Delta _{\vec{k}}^{s}(T)^{2}}{%
dT}+\frac{d\Delta _{\vec{k}}^{d}(T)^{2}}{dT}+2\cos (\theta )[\frac{d\Delta _{%
\vec{k}}^{s}(T)}{dT}\ \Delta _{\vec{k}}^{d}(T)+\Delta _{\vec{k}}^{s}(T)\frac{%
d\Delta _{\vec{k}}^{d}(T)}{dT}]  \label{dergap}
\end{eqnarray}
Following Eq.(\ref{gaptemp}) and irrespective of the gap symmetry, the
derivative of $\Delta (T)$ and $\Delta ^{2}(T)$ with respect to the
temperature are given by: 
\begin{eqnarray}
\frac{d\Delta (T)}{dT} &=&\left\{ 
\begin{array}{ll}
-\frac{\textstyle{T_{c}}}{\textstyle{T^{2}}}\ \sqrt{\frac{\textstyle{T}}{%
\textstyle{T_{c}-T}}}\ \frac{\textstyle{\Delta (0)}}{\textstyle{\cosh ^{2}[%
\sqrt{\frac{\textstyle{\ T_{c}-T}}{\textstyle{T}}}]}}\ \ ;\  & 
\mbox{ if T $\neq$
$T_c$} \\ 
-\infty \ \ \ \ \ \ \ \ \ \ \ \ \ \ \ \ \ \ \ \ \ \ \ \ \ \ \ \ \ \ \ \ \ \
\ \ \ \ \ \ \ \ \ \ \ \ \ \ \ \ \ \ \ \ \ \ \ \ \ \ \ \ \ \ \ \ \ \ \ \ \ \
\ \ \ \ \ \ \ ;\  & \mbox{if $T=T_c$,}
\end{array}
\right.  \label{dertg} \\
\frac{d\Delta ^{2}(T)}{dT} &=&\left\{ 
\begin{array}{ll}
-\frac{\textstyle{2T_{c}}}{\textstyle{T^{2}}}\ \sqrt{\frac{\textstyle{T}}{%
\textstyle{T_{c}-T}}}\ \frac{\textstyle{\Delta ^{2}(0)\ \tanh [2\sqrt{\frac{%
\textstyle{T_{c}-T}}{\textstyle{T}}}]}}{\textstyle{\ \cosh ^{2}[\sqrt{\frac{%
\textstyle{T_{c}-T}}{\textstyle{T}}}]}}\ \ ; & \mbox{ if T $\neq$ $T_c$} \\ 
-\ \frac{\textstyle{2\Delta ^{2}(0)}}{\textstyle{T_{c}}}\ \ \ \ \ \ \ \ \ \
\ \ \ \ \ \ \ \ \ \ \ \ \ \ \ \ \ \ \ \ \ \ \ \ \ \ \ \ \ \ \ \ \ \ \ \ \ \
\ \ \ \ \ \ \ \ \ \ \ \ \ \ \ \ \ \ \ \ ; & \mbox{if
$T=T_c$,}
\end{array}
\right.  \label{dertg2}
\end{eqnarray}
As can be seen from Eqs.(\ref{dergap}) and (\ref{dertg}), in the case of $%
s+d $ mixing an square root singularity in the specific heat is obtained if $%
T_{c}^{s}$ and $T_{c}^{d}$ are different. Otherwise, a smooth
dependence on $T/T_{c}$ is achieved. However, up to now experimental reports
on the presence of this kind of singularity are still absent. Hence, in
order to observe the $s+d$ symmetry the critical temperatures for the both
components ($\Delta _{s}$ and $\Delta _{d}$) must be the same. It is
important to remark that this result have nothing to do with the EDOS or any
criteria of stability for the phase under certain crystalline symmetry. For
the $s+id$ mixing the square root singularity does not appear, because the
relative phase between the two components of the gap is equal to $\pi /2$
and the cosine function in Eq.(\ref{dergap}) is zero. Further we shall deal
with the influence of the $s+id$ mixing on $C_{es}(T)$ and $H_{c}(T)$.

Transforming the sum into integrals over the first Brillouin zone and
applying similar methods as in Ref. \cite{liu}, $C_{es}$ can be cast into: 
\begin{eqnarray}
C_{es}(T) &=&\frac{4R}{\sqrt{1-\alpha ^{2}}}\ (\frac{2t}{k_{B}T_{c}})^{2}(%
\frac{T_{c}}{T})^{2}\ \int_{-\omega _{c}}^{\omega _{c}}\ d\xi \
\int_{v_{l}}^{v_{u}}\ dv\ \frac{N(v,\xi )}{W(v,\xi )}f(W)(1-f(W))  \nonumber
\\
&&[W^{2}(v,\xi )-\frac{1}{2}(\frac{k_{B}T_{c}}{2t})^{2}T\frac{d}{dT}[(\frac{%
\Delta _{s}}{k_{B}T_{c}})^{2}+(\frac{\Delta _{d}}{k_{B}T_{c}}%
)^{2}D^{2}(v,\xi )]],
\end{eqnarray}
where 
\begin{eqnarray}
N(v,\xi ) &=&\frac{1}{2\pi ^{2}}\ \frac{1}{\sqrt{(1-v)(1+v)(v-v1)(v2-v)}}, 
\nonumber \\
W^{2}(v,\xi ) &=&\xi ^{2}+(\frac{k_{B}T_{c}}{2t})^{2}[(\frac{T_{c}^{s}}{T_{c}%
})^{2}\ (\frac{\Delta _{s}}{k_{B}T_{c}^{s}})^{2}+(\frac{T_{c}^{d}}{T_{c}}%
)^{2}\ (\frac{\Delta _{d}}{k_{B}T_{c}^{d}})^{2}D^{2}(v,\xi )],  \nonumber \\
D(v,\xi ) &=&-\ (\xi +2v)\ \,\ \ v1=-\ (1+\xi ),\ \ v2=(1-\xi ),  \nonumber
\end{eqnarray}
$R$ is the universal gas constant and $\omega _{c}=\varepsilon _{c}/2t$ is
the dimensionless energy cut-off. In the following for $2\Delta
_{s}(0)/k_{B}T_{c}^{s}$ and $2\Delta _{d}(0)/k_{B}T_{c}^{d}$ the values of 4 
\cite{davidq} and 4.26 \cite{tsuei3} have been used, respectively. The
integration limits over $v$ (lower $v_{l}$ and upper $v_{u}$) are defined as
follows: for $-\omega _{c}<\xi <0$, $v_{l}=v1$ and $v_{u}=1$, while for $%
0<\xi <\omega _{c}$, $v_{l}=-1$ and $v_{u}=v2$. For numerical evaluation,
the values $\omega _{c}=0.07$ ($\varepsilon _{c}=35meV$) and $%
k_{B}T_{c}/2t=0.017$ ($T_{c}=100K$) have been used. The function $N(\xi ,v)$
is a measure of the EDOS for the respective symmetries of the gap as it can
be seen by the following integrals \cite{liu}: 
\begin{eqnarray}
N_{d}(\xi ) &=&\int_{v_{i}}^{v_{s}}\ dvN(v,\xi )\ D^{2}(v,\xi ), \\
N_{s}(\xi ) &=&\int_{v_{i}}^{v_{s}}\ dvN(v,\xi ).
\end{eqnarray}
$N_{s}(\xi )$ coincides with the exact EDOS of the tight binding model on
the square lattice, given by a complete elliptic integral of the first kind
(see Ref.\cite{davidq}) and $N_{d}(\xi )$ could be regarded as an effective
EDOS for the d-wave pairing.

The dependence of $C_{es}/C_{en}$ ($C_{en}$ is the specific heat in the
normal state) on the reduced temperature $T/T_{c}$ is presented in Fig.2 for 
$c=0$ and $c=0.96$. The first case corresponds to a nearly pure s-wave while
the second has a d-wave character. As can be seen from this picture the jump
due to the s-wave gap at $T_{c}$ is smaller than that of the d-wave one.
This is in contrast with the conventional anisotropic superconductors, where
the specific heat jump in the isotropic case is larger than that for the
anisotropic one \cite{clem}. The above fact is explained as follows. Taking
into account Eqs.(\ref{gap-s}), (\ref{gap-d}), and (\ref{gaptemp}) the
specific heat jump can be written as: 
\[
\Delta C(T_{c})=\left\{ 
\begin{array}{ll}
18.14k_{B}\ \int_{-2}^{2}\ d\xi \ N_{d}(\xi )\ f(|\xi |)\ [1-f(|\xi |)]\ \ \ 
& ;\ \ \ \mbox{d-wave} \\ 
&  \\ 
16k_{B}\ \int_{-2}^{2}\ d\xi \ N_{s}(\xi )\ f(|\xi |)\ [1-f(|\xi |)]\ \ \  & 
;\ \ \ \mbox{s-wave}
\end{array}
\right. 
\]
We note that the term $f\ (1-f)$ is the same for both symmetries. The square
of the cosine difference appearing in the EDOS $N_{d}(\xi )$ for the d-wave
case plays a key role for the understanding of the difference of $\Delta
C(T_{c})$ between the two symmetries. It is important to note that $%
N_{d}(\xi )$ function has a stronger singularity than that coming from the
s-wave gap explaining why the $\Delta C(T_{c})$ value for d-waves is larger
than the corresponding to the s-symmetry. In the inset a comparison between
the s- and d-wave specific heats at low temperature is shown.

The evolution in term of the driven parameter $c$ for the temperature
dependence of $C_{es}(T)/C_{en}(T_{c})$ is shown in Fig.3. For $c\leq 0.385$
the s-wave gap is dominant (see Fig.3(a)), while for $c>0.385$ the $T_{c}$
is ruled out by the d-wave (see Fig.3(b)). The double peak transition is
better seen for $0.37\leq c\leq 0.40$. In Fig.3(a) (Fig.3(b)) the peaks
numbered by 1 to 4 (1 to 3) correspond to the d-wave (s-wave) contribution.
The results depicted in Figs.3(a) and (b) constitute a direct consequence of
the difference in the jump $\Delta C(T_{c})$ for the mixed symmetries. These
results are qualitatively in good agreement with those reported in \cite
{choy,butera,junod} where a double peak transition has been observed. It is
worth noticing that in recent measurements on $Y_{0.8}Ca_{0.2}Ba_{2}Cu_{3}O_{7-%
\delta }$ \cite{loram} this structure apparently seems to be absent.
Nevertheless, these data show, for certain values of Oxygen concentration, a
small feature that could be interpreted in the framework of the mixed waves
scenario.

The behavior of $C_{es}(T)/C_{en}(T_{c})$ at very low temperatures is
depicted in Fig.4 for values of $c>0.4$, where the d-wave contribution is
dominant. The quadratic behavior observed in a wide range of temperatures
for the d-wave case is in contrast with the linear dependence obtained in
Ref. \cite{dorbolo}. This quadratic law is consistent with the experimental
measurements of Ref.\cite{momono} and the phenomenological approach which
assumes a linear dependence on energy for the EDOS in the superconducting
state. As soon as $c$ reaches the values 0.52, in the low temperature region 
$C_{es}(T)/C_{en}(T_{c})$ goes through a small and broad shoulder around $%
T=0.005T_{c}$ superposed onto the quadratic behavior law. This anomaly seems
to be related with the concrete topology of the gap directly reflected in
the lines of constant quasiparticle energy $E_{\vec{k}}$ of the d-wave
superconductor. It appears only when the van Hove and the d-wave gap are
simultaneously present, otherwise they can not appear.

The thermodynamic critical magnetic field as a function of temperature can
be obtained from the usual BCS expression \cite{schri1}. In Fig.5 the
dependence of the reduced critical magnetic field $H_{c}(T)/H_{c}(0)$ on the
reduced temperature $T/T_{c}$ is shown for different values of $c$
parameter. As can be seen see in Fig.5(a), for $c\le 0.39$ the s-wave
component is dominant and the phase transition to a single symmetry state is
clearly observed. Here, the slope at $T_{c}$ associated with the s-wave gap
is lower than that coming from the conventional BCS. Moreover, when the
d-wave component is dominant the phase transition is almost unobservable and
the temperature dependence in this case shows an small departure from the
old parabolic law (compare Figs.5(a) and (b)). This quasi-parabolic law in
the d-wave case is a direct consequence of the zeros in the quasiparticle
excitation energy, which also provides a power law in the low temperature
regime of the specific heat (see inset in Fig.2).

\section{Conclusions.}

A toy model within a van Hove BCS scenario which takes into account the
existence of mixing of gap symmetries has been developed. Within this
framework the specific heat dependence on the temperature was computed
resulting in a double peak transition whose resolution is very sensitive to
the amount of the d-wave component present in the gap. For a pure d-wave
gap, a quadratic law at low temperatures with a superposed shoulder at $%
T=0.005T_{c}$ is obtained. On the other hand, $\Delta C$ jump at $T=T_{c}$
of the pure d-wave component is larger than that coming from an s-wave one.
These features are due to the joint effect of the gap anisotropy and van
Hove singularity. It is expected that the influence of anisotropy will be
reinforced in those directions where it is allocated. Our calculations show
that the critical magnetic field follows a quasi-parabolic law with
temperature which results from the isolated zeros in the quasiparticle
excitation energy. The anisotropy effect along with the van Hove singularity
establish a qualitative difference with respect to old conventional
superconductors where a constant EDOS is considered. Moreover, it seems that
a model with a dominant d-wave contribution plus an small s-wave component
may show a closes fitting with the reported experimental measurements.

\newpage

\begin{center}
{\Large {\bf Figure captions}}
\end{center}

\begin{enumerate}
\item[Fig.1]  Lines of constant quasiparticle excitation energy $E_{\vec{k}}$
in unit of $2t$ as a function of the dimensionless wave vector components $%
X=k_{x}a$ and $Y=k_{y}a$ for a gap with a pure s-wave (upper left corner),
and d-wave (upper right corner) symmetries, an $s+d$ (lower left corner) and 
$s+id$ (lower right corner) mixing.

\item[Fig.2]  Electronic specific heat in the superconducting state in units
of $C_{en}(T_{c})$ as a function of the reduced temperature $T/T_{c}$ for
the values of $c=0$ (s-wave) and $c=0.96$ (d-wave gap). The inset shows the
range of low temperatures.

\item[Fig.3]  Electronic specific heat $C_{es}(T)/C_{en}(T_{c})$ as a
function of $T/T_{c}$ for different values of the driven parameter {\bf {\it %
c}} in the region of gap symmetry mixing: (a) $c=0.37$ (1), 0.375 (2), 0.38
(3), and 0.385 (4); (b) $c=0.39$ (1), 0.395 (2), and 0.4 (3).

\item[Fig.4]  Very low temperature behavior of the superconducting specific
heat for different values of $c$: 0.43 (1), 0.46 (2), 0.49 (3), 0.52 (4),
0.55 (5). A superposed shoulder is clear at $T=0.005T_{c}$ for values of $c$
equal to 0.52 and 0.55.

\item[Fig.5]  Thermodynamic critical magnetic field in units of $H_{c}(0)$
as a function of the reduced temperature $T/T_{c}$ for different values of
the driven parameter {\bf {\it c}} in the region of gap symmetry mixing: (a) 
$c=0.37$ (1), 0.375 (2), 0.38 (3), and 0.385 (4); (b) $c=0.39$ (1), 0.395
(2), and 0.4 (3) (see text for explanation).
\end{enumerate}

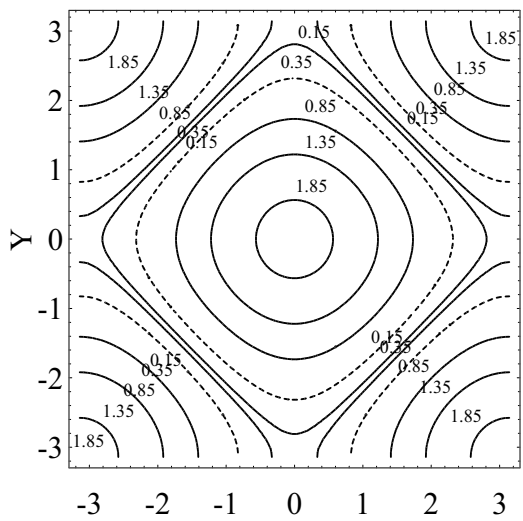
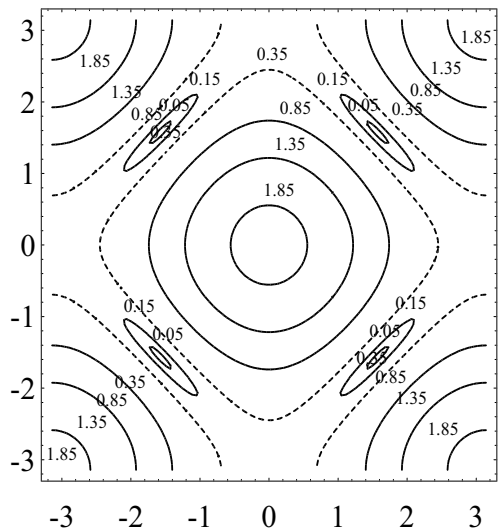
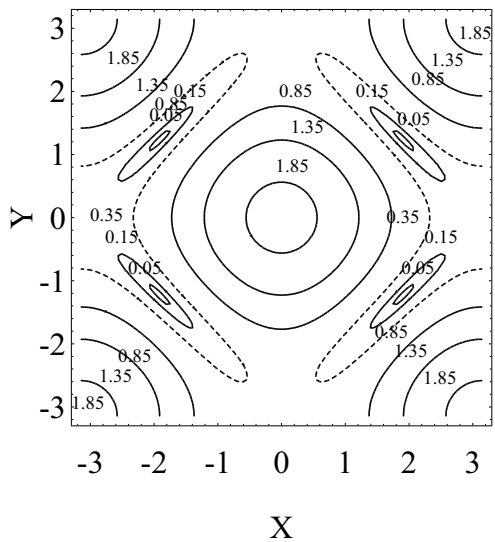
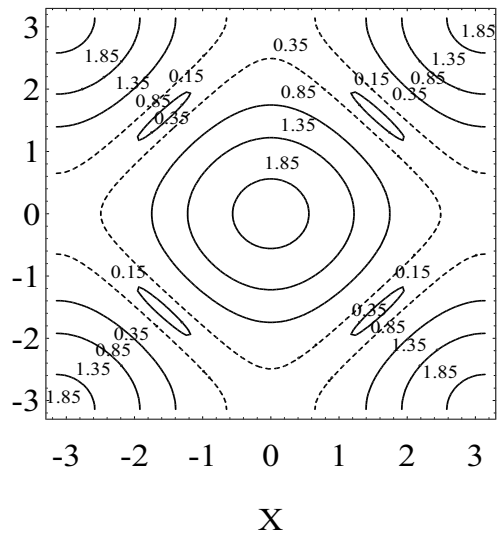
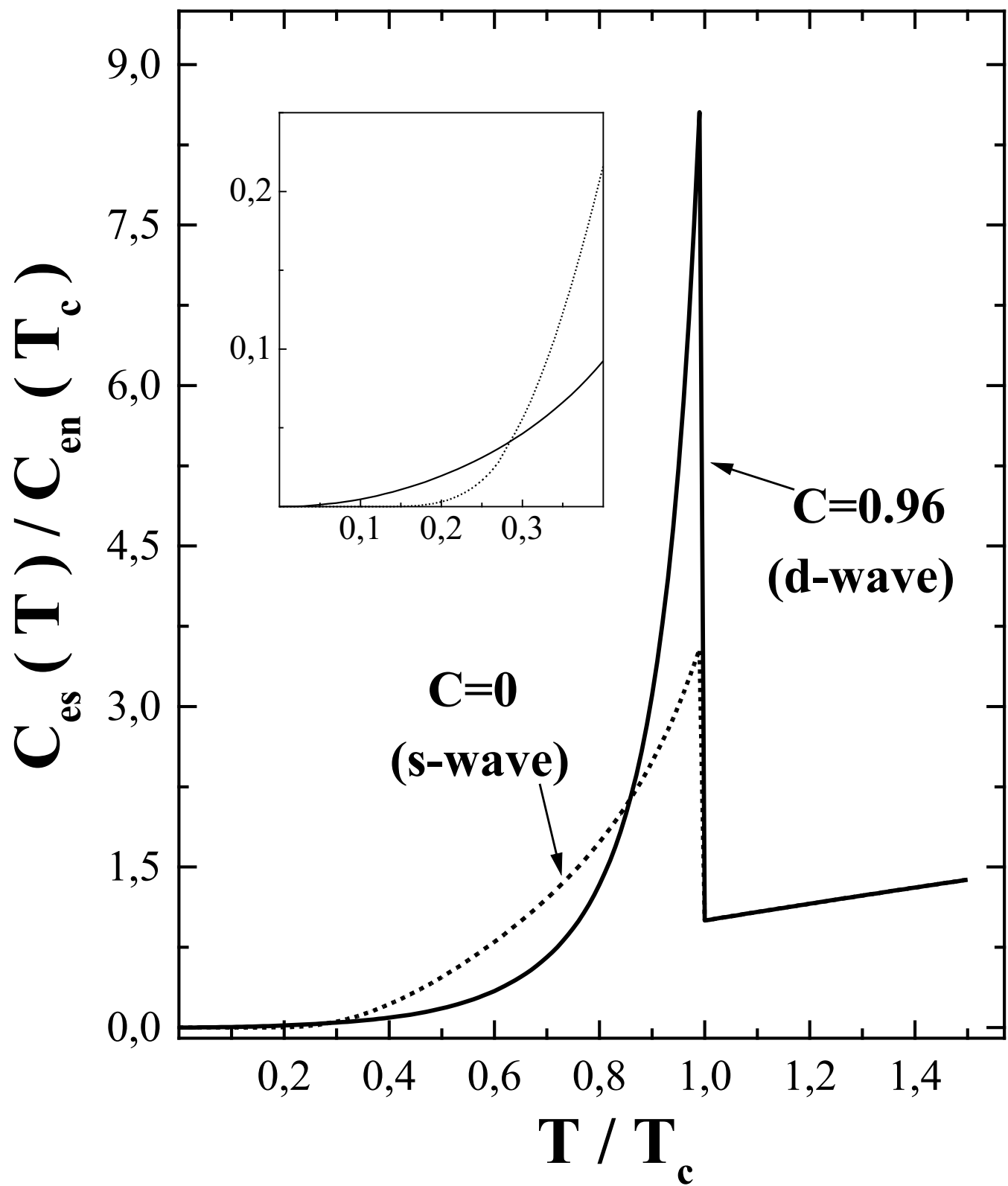
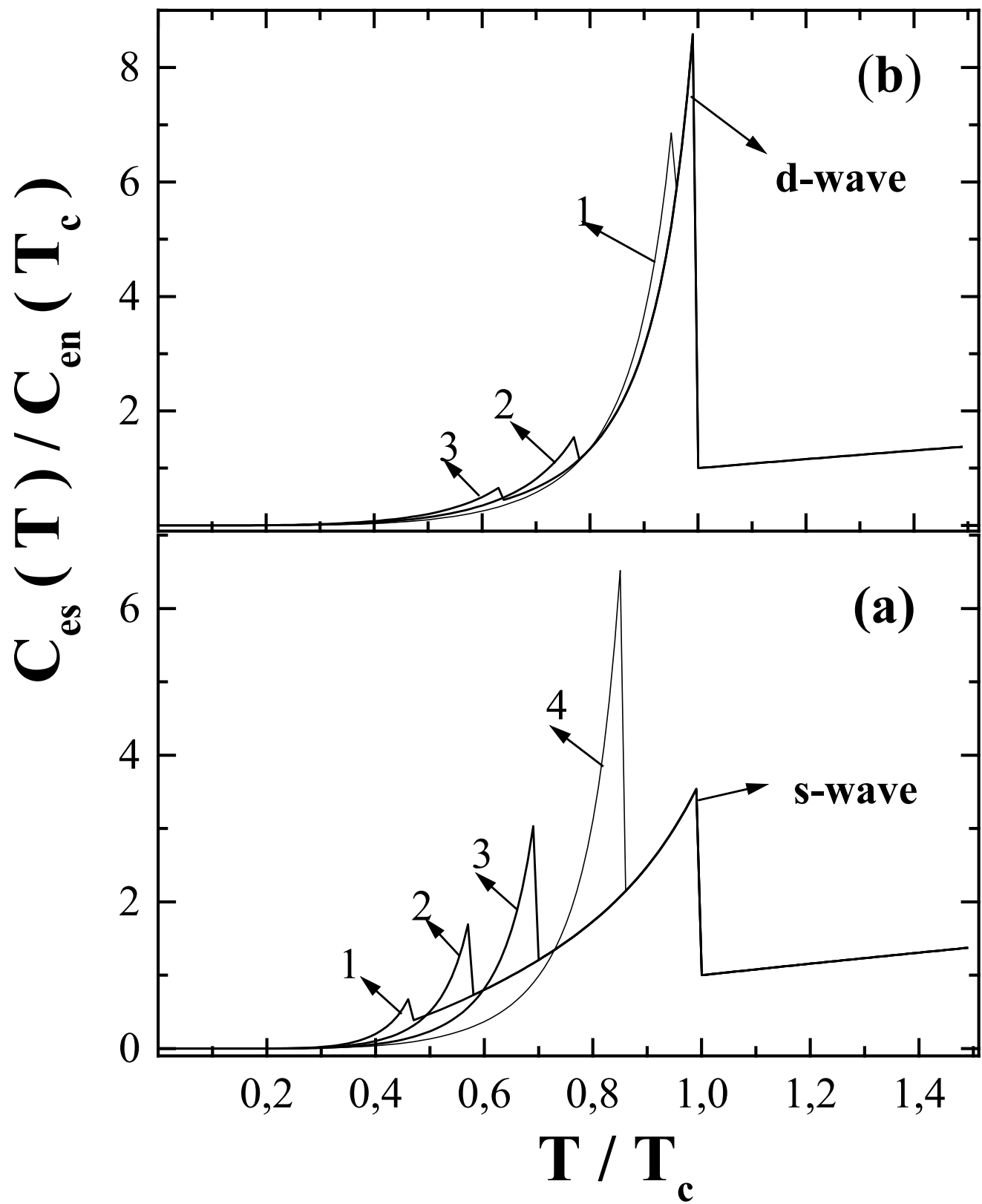
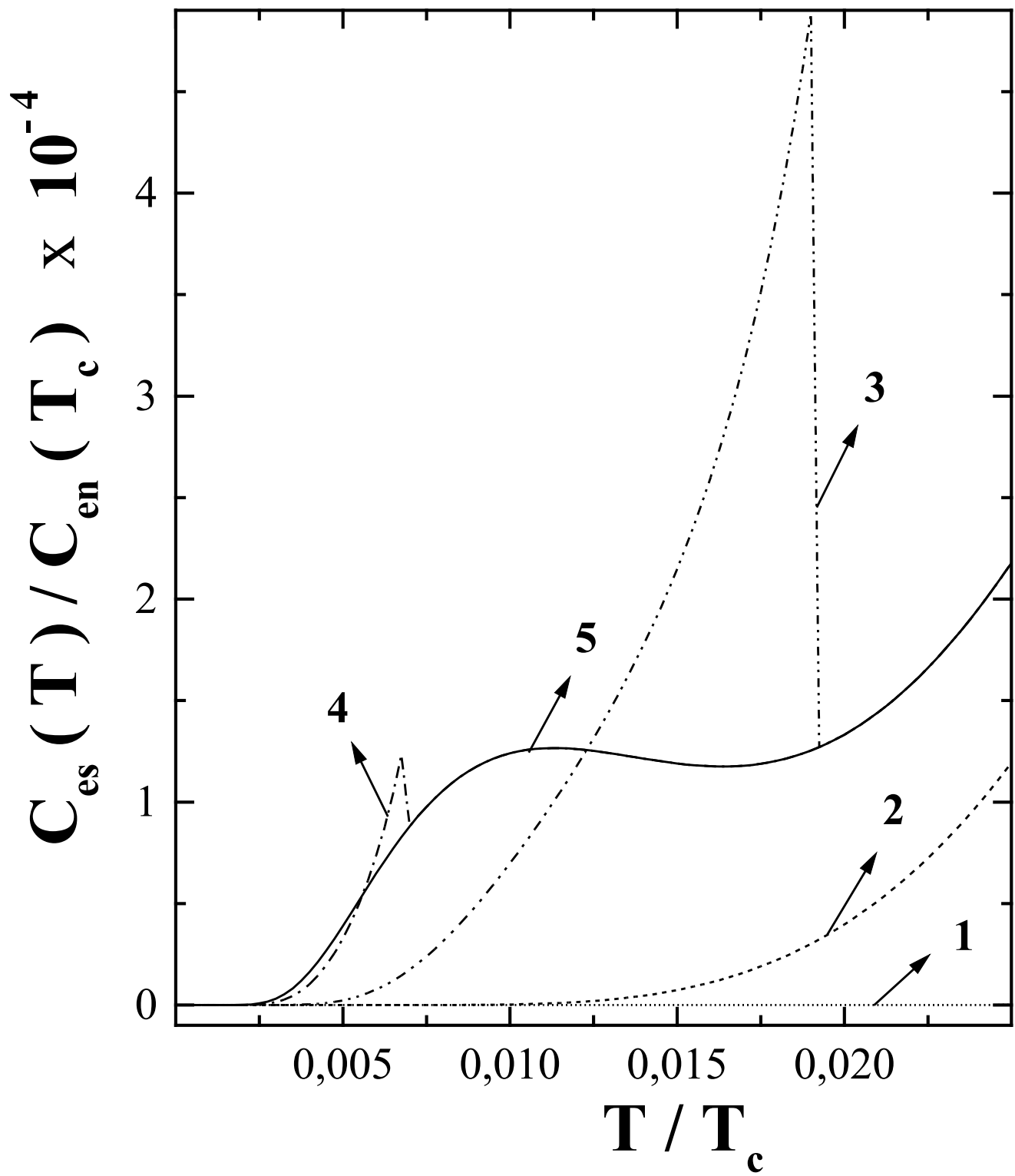
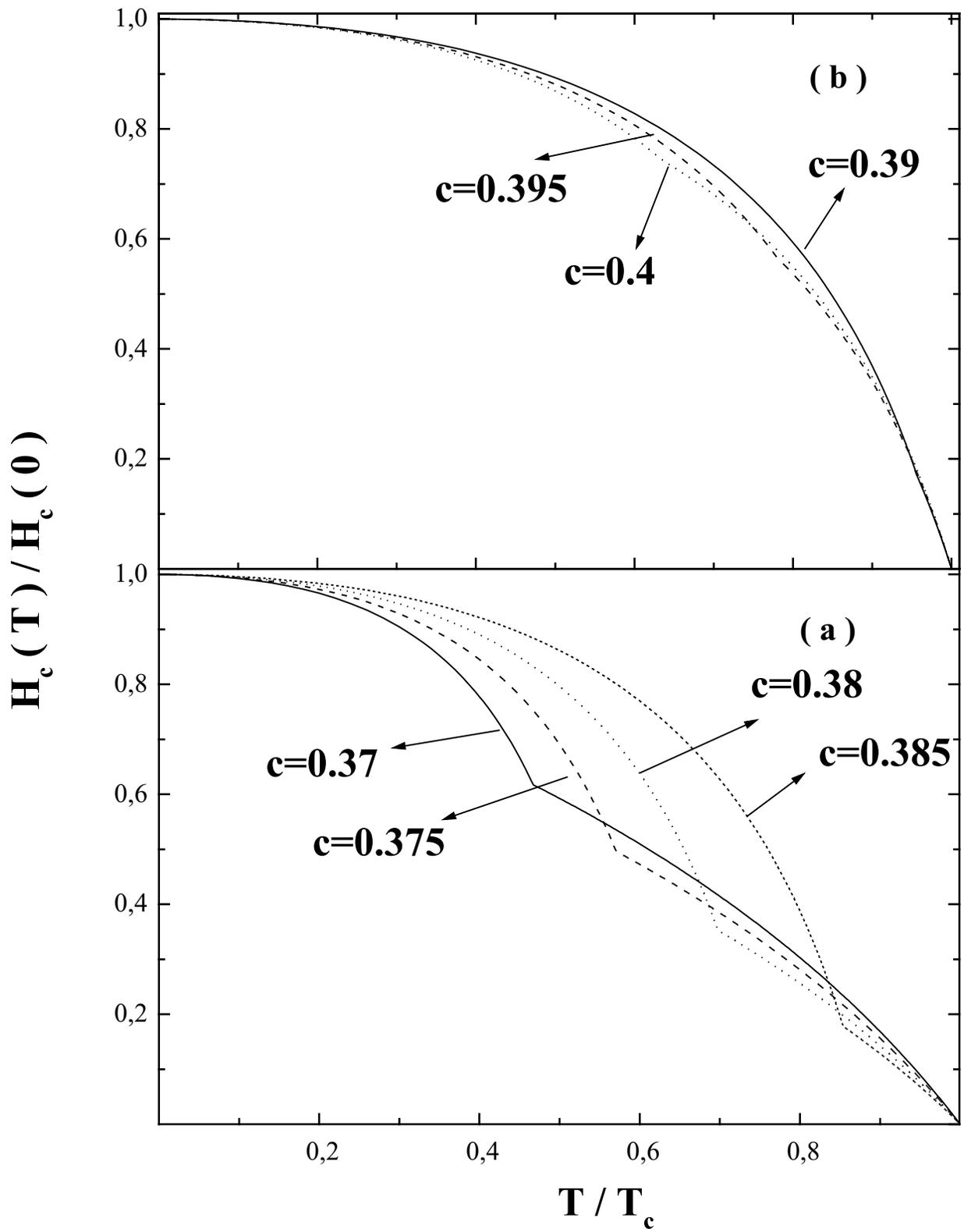
\end{document}